# Non-equilibrium thermodynamics and collective vibrational modes of liquid water in an inhomogeneous electric field

Adam D. Wexler[a*], Sandra Drusová[a], Jakob Woisetschläger[b], Elmar C. Fuchs[a]

[a]*Applied Water Physics, Wetsus European Center of Excellence for Sustainable Water Technology, 8911MA Leeuwarden, Netherlands*

[b]*Institute for Thermal Turbomachinery and Machine Dynamics, Working Group Metrology - Laser Optical Metrology, Technical University of Graz, 8010 Graz, Austria*

* corresponding author: adam.wexler@wetsus.nl +31 (0)58 284 3013 http://www.wetsus.eu

**Abstract**

In this experiment liquid water is subject to an inhomogeneous electric field ($\nabla^2 E_a \approx 10^{10}\ V/m^2$) using a high voltage (20 kV) point-plane electrode system. With interferometry it was found that the application of a strong electric field gradient to water generates local changes in the refractive index of the liquid, polarizes the surface and creates a downward moving electro-convective jet. A maximum temperature difference of 1 °C is measured in the immediate vicinity of the point electrode. Raman spectroscopy on water reveals an enhancement of the vibrational collective modes (3250 cm$^{-1}$) as well as an increase in the local mode (3490 cm$^{-1}$) energy. This bimodal enhancement indicates the spectral changes are not due to temperature. The intense field gradient thus establishes an excited subpopulation of vibrational oscillators far from thermal equilibrium. Delocalization of the collective vibrational mode spatially expands this excited population beyond the microscale. Hindered rotational freedom due to electric field pinning of molecular dipoles retards heat flow and generates a chemical potential gradient. These changes are responsible for the observed changes in refractive index and temperature. It is demonstrated that polar liquids can thus support local non-equilibrium thermodynamic transient states critical to biochemical and environmental processes.

1. Introduction

This experiment using a high voltage (20 kV) point plane electrode system was designed to better understand the interaction between water and high electric fields during the so-called "floating water bridge" experiment without electric current interfering. The floating water bridge is a special case of an electrohydrodynamic (EHD) liquid bridge. It forms when a high potential difference (kV cm$^{-1}$) is applied between two beakers of water pulling water up to the edges of the beakers and forming a free hanging water string through air connecting the two beakers. The discovery of this phenomenon dates to the late 19th century, when in 1893 Sir William Armstrong first reported this experiment.[1] While the macroscopic EHD processes are well understood [2,3] microscopic dynamics still need more research.

The interaction of polar liquids with moderately strong fields (10$^6$ V m$^{-1}$) has been studied extensively within the field of electrohydrodynamics and are commonly employed in the production of molecular beams [4,5], inkjet printing[6], electrospinning [7], and desalination.[8] EHD liquid bridges have recently provided unique insights into the non-equilibrium molecular physics of polar liquids such as water because they are both large and stable enough to use with experiments that access the molecular length scale but require a macroscopic sample volume – e.g. radiation scattering [9–13], and spectroscopy.[14,15] A previous Raman study on vertical EHD bridges found a small fraction of molecules in the bridge are polarized,[16] presumably in association with charge carriers, which in the case of water are predominantly solvated protons.[17] The typical current measured in bridges, whether driven by direct or alternating fields, is on the order of a few milli-Amperes. Another characteristic of EHD bridges is the presence of strong inhomogeneous electric field gradients ($\nabla^2 E_a \approx 10^{10}\ V/m^2$) where the bulk liquid transitions to the bridge.[18,19] It is interesting to consider what effect such a field gradient alone, without the presence of solvated charge carriers, may have on molecular polarization. It is noteworthy that bridges are only stable in the presence of the field, as soon as the field is switched off, the floating liquid bridge ruptures under the influence of surface forces and gravity – the electric field is an essential stabilizing force.

Dielectric relaxation in polar liquids can exhibit marked deviations from theoretical expectations. This is often attributed to microscopic polarization dynamics and molecular coupling interactions.[20] The study of dielectric relaxation is typically conducted as a bulk measurement over frequencies from mHz to GHz. Advances in optical spectroscopy have extended the range of this technique to the THz, though all optical interactions with matter depend upon the interaction of the time varying electric field and the electrical response in the material (i.e. refractive index). The applied electric field is typically that supplied by the probing radiation. At low frequencies an offset or bias voltage can also be applied typically in the range of a few tens of volts but can range up to a few kV. For a typical dielectric spectroscopy the applied field is on the order of $10^{-5}$ V nm$^{-1}$. Experimentally, the electric field is assumed to approach the homogeneous limit as electrodes, and probes are designed to minimize the fringing field.[21] In case of molecular probes the number density is kept low enough that the solution may be considered in the isotropic limit.[22] Dielectric breakdown depends on the dielectric properties of the material as well as the parameters of the electric field. At present the engineering of high power electrical systems must rely upon empirical design rules and safety margins to avert such cataclysmic events.[23] Our understanding of the processes governing the interaction of liquid matter with electric fields are not complete – largely because a clear understanding of long range and many-body interactions in liquid matter evades us.[24]

Further developing the theoretical framework of non-equilibrium dielectric processes can improve our understanding of biological and natural systems. The dielectric response of water, aqueous solutions, and biochemical machinery is critical to the functioning of cellular processes.[25] Electric fields are a hallmark of living systems [26] and like those found in the environment [27,28] they are typically inhomogeneous again with field strengths on the order of $10^6$ V m$^{-1}$ ($10^{-3}$ V nm$^{-1}$). Thus in order to better understand the connection between dielectric processes, the molecular environment, and inhomogeneous electric fields interferometry and infrared Raman spectroscopy were combined to interrogate the response of water in a point-plane electrode system.

## 2. Experimental

*2.1 General experimental construction*

The inhomogeneous electric field was produced using a point-plane electrode geometry and is illustrated in figure 1, panel a. Type I HPLC grade purified water (18.2 MΩ, Cat No. ZMQSP0D01, Millipore Corp., MA, USA) was filled into a glass cuvette (52 mm x52 mm x 14mm, Cat. No. 704.003, Hellma analytics, Müllheim, Germany), and this was then placed in between the electrodes. The planar electrode is made from copper foil (Cat. No. 1181 3M, MN, USA) which is embedded in an insulating PMMA block (100 mm x 100 mm x 25 mm) and connected to the high voltage output of a DC power supply (HCP 30000-300, FuG Elektronik GmbH, Schechen, Germany). The PMMA block had a 5 mm deep groove machined into the top where the cuvette was placed into direct contact with the upper electrode surface. The planar electrode had nearly the same dimensions (9.5 mm x 50 mm x 2.6 mm) as the bottom of the cuvette. A stainless steel acupuncture needle (SJ.25x40, 40 mm x 0.25 mm, Seirin Corp., Shizuoka, Japan) was used to form the point electrode and connected to the ground terminal of the high voltage source. The electrode was affixed to a rigid insulating armature constructed from polypropylene. The needle was bent at a right angle ~5 mm beyond the handle in a manner so as to prevent the electrode from blocking the incident laser beam. For each experiment the electric potential was set to 20 kV and applied as a step function. During operation a small leakage current <10 µA could be measured and was presumably due to corona discharge. The electric field for this configuration was modeled using COMSOL Multiphysics (v. 5.0 Comsol, Inc., Stockholm, Sweden) and is shown in figure 1b-c.

Twenty milliliters (20 mL) of water were used for each measurement. This brought the liquid surface to a height of 40mm from the bottom of the cuvette and to within 1mm from the tip of the needle. The water for the interferometric investigations was kept in borosilicate glass bottles heated in a water bath to 70°C and transferred to the unheated cuvette just prior to energizing the experiment and the water replaced for every new measurement. The elevated temperature enhanced the polarization response of water and the resulting relative shift in refractive index meant changes were better resolved by the interferometer. Raman measurements were conducted using room temperature water (20±1°C). Temperature was measured during the experiments using a fiber optic probe (OTG-F, ProSens, OpSens Inc., Quebec, Canada) which could be placed at several locations in the experiment including directly beneath the point electrode. The sensor system spectroscopically monitors the band gap energy of a GaAs crystal to measure temperature and as such is not affected by the applied electric field nor any field induced changes in the sample. This probe has a maximum resolution of 0.01°C and readout speed up to one kHZ repetition rate. During interferometric measurement the most significant response was observed in less than 10 seconds so the overall temperature change in the liquid was negligible relative to the field effect.

*2.2 Interferometry*

A Mach-Zehnder interferometer was constructed with 2 meter base length and is diagramed in figure 2. The source laser was a frequency stabilized continuous wave single-line mode diode laser (DL640-050-S, P=50mW, λ=635±2 nm, 1.1 mm Gaussian beam diameter, 1 mrad divergence, Crystalaser, NV, USA) with built in optical isolator and polarization ratio 100:1, a half wave plate was positioned to rotate the polarization direction in the interferometer. The beam was spatially filtered using a 20x microscope objective and 5µm pinhole, subsequently over expanded with a biconcave lens and finally collimated using a large diameter plano-convex lens. The over-expanded beam was blocked with a square aperture placed at the entry port of the first beam splitter cube. This assured that the most central portion of the Gaussian beam filled the full aperture and provided good uniformity of illumination throughout the test subject. The mirrors in the interferometer section were inclined slightly to produce a carrier fringe system that allowed the disambiguation of the phase shift in the recovered interferograms. The cuvette test section was position in the interferometer in direct view of the recording camera (Photron FASTCAM SA1.1) fitted with an 18-270 mm, F/3.5-6.3 zoom lens (Cat. No. B008,Tamron, NY, USA). Images were collected at 50 frames per second full frame resolution (1024x1024 pixels). The recorded images were evaluated using IDEA software (http://optics.tugraz.at/) using a fast Fourier transform (FFT) based digital fringe evaluation.[29] For the interferometry studies the time-resolved response of the sample was recorded as the voltage was applied according to a rectangular step function. The voltage profile was to step between 0 and 20 kV with a rising edge <500 ms, a stable plateau between 10 – 60 seconds and a falling edge which fully discharged the experiment back to ground potential in under 2 s.

*2.3 Mid-Infrared Raman Imaging Spectroscopy*

The mid-infrared (mid-IR) vibrational modes of water are Raman active and can be accessed using visible lasers of high power which have better penetration depth than direct mid-IR spectroscopy. This is ideal for probing the spatial variation in the vibrational modes of water under electrical stress. The cuvette point-plane electrode system previously described was arranged as shown in figure 3. A collimated and polarization controlled beam from an argon ion laser (Innova 6W Ar+, P=400mW, 514.5 nm, 1.5 mm beam diameter, 0.5 mrad divergence, Coherent, USA) was guided into the cuvette either perpendicular or parallel to the liquid surface. The beam delivery optics provide a very uniform and tight focus over several centimeters, given the beam characteristics stated above, the focal spot size is 150 µm and the confocal parameter is 58.89 mm which is broader than the entire sample cuvette. Thus it is assured that within the interrogated spectral imaging region the illumination power and beam shape are uniform. Furthermore, care was taken to assure that the beam and acupuncture needle did not intersect, the laser beam always passing in front of the needle in those cases where the needle tip was placed deeper than the laser beam. For water at ambient conditions the extinction coefficient for water at 515nm is $3.96 \times 10^{-4}$ cm$^{-1}$, note unit is in terms of propagation distance and not wavenumbers.

The incident electric field polarization ($E_0$) of the laser beam was controlled by a zero-order half wave plate. When the beam was parallel to the liquid surface $E_0$ was also parallel to the central axis of the applied static electric field gradient ($E_a$) imposed by the needle electrode. When $E_0$ was perpendicular to the liquid surface it was also perpendicular to $E_a$. In order to maintain the imaging condition the entrance slit and laser beam direction were kept parallel. This effectively meant turning the spectrograph on its side when the beam was parallel to the liquid surface. This also reduced systematic error due to polarization dependent differences in the dispersion efficiency of the spectrograph gratings as the relative orientation of the grating rulings and $E_0$ was maintained throughout the experiments. There are a number of electric field vectors to keep track of in the experiments considered here. For simplicity we will only use the vertical (V) and horizontal (H) directions in discussing each. The first field is that of the voltage applied to the point plane electrodes, while the real field is a gradient with complex shape (see figures 1b and 1c) the needle geometry defines the principal axis of the field ($E_a$) as V polarized. The electric field of the laser beam ($E_0$) must be orthogonal to the beam propagation Poynting vector and was rotated by the λ/2 plate to be parallel to the long axis of the cuvette. Thus when the beam was parallel to the liquid surface $E_0$ was V polarized, and when the beam was perpendicular $E_0$ was H polarized. The electric field of the scattered light ($E_s$) was analyzed in both the H and V polarizations. For each experimental configuration four sets of spectra were recorded, two VV (or HH) orientations and two VH (or HV) orientations depending on the beam Poynting vector.

The Stokes shifted Raman scattered light was collected using a photographic lens (Mamiya Secor SX135 mm F 2.8, Mamiya, Japan) positioned to image the fluid region nearest the tip of the needle onto the entrance slit of a spectrograph (Acton SpectraPro 2300i, 600 lines/mm grating, 30 μm entrance slit, resolution 3.9 cm$^{-1}$, Acton, NJ, USA). Prior to the vertical entrance slit the light passed through a linear polarizing filter that could be rotated so the transmitted electric field vector was either parallel or perpendicular to the normal of the electric field gradient. The monochromator produced spectrally dispersed light which was imaged using a cooled, intensified CCD camera (NanoStar, LaVision, Göttingen, Germany). The integration time for each collected frame was 100 milliseconds with an intensifier gain of 50. For each experimental configuration 50 frames were collected and summed together. Each frame contained 512 spectra with 640 spectral points each. The sum of the background intensity images recorded with the entrance slit blocked was subsequently subtracted from the experimental images. This effectively removed signal contributions due to thermal, electronic readout, and shot noise and improved the resulting image contrast. All image recording and post-processing was carried out using Lavision DaVis 7.2.1.64 software (Lavision, Göttingen,

Germany). The applied electric field intensity was either zero or 20 kV. Voltage was again applied as a step function however, now the system was allowed to equilibrate for at least one minute prior to recording spectra.

## 3. Results and Discussion

*3.1 Interferometry*

In an interferometer the light path is split at the first beam splitter into a reference and test arm, the latter passes through the test section, in this case the cuvette, before both beams are recombined in the second beam splitter. The light passing through the test section suffers a phase lag which is dependent upon the refractive index of the material. Depending upon the phase lag the light will interfere constructively or destructively and creates bright and dark bands or interference fringes. The measurement is line-of-sight, meaning that it is only sensitive to light whose propagation direction (i.e. pointing vector) is towards the detector, off axis or scattered radiation is not detected, and the phase lag is integrated over the path in the cuvette. This makes interferometry quite sensitive to local perturbations in the complex dielectric permittivity (i.e. refractive index) of the material under study and provides a measure of the polarizability as well as the dipole number density in the material.

The recorded interferograms clearly show that application of a denseinhomogeneous electric field of the magnitude used in these experiments ( $\nabla^2 E_a \approx 10^{10}\ V/m^2$) will generate polarization forces which disturb not only the liquid nearest the point electrode but generate perturbations along the entire air-liquid surface and deep into the bulk. The processed results from the interferometric investigations are shown in figure 4. A representative image from a raw interferogram showing the placement of the two electrodes as well as the evaluated region of interest (ROI) is displayed in panel (4a). The evaluated data panels (4b) to (4g) with correction for phase ambiguity (i.e. positive phase shift for increase in density) show that there are several responses of the liquid and that the time evolution of each feature is different. Two major responses will be discussed: surface polarization and downward jets. The application of the electric field results in a strong bending of the fringes in the top few millimeters of liquid and likewise the removal of the field causes the fringes to return to their original position. The magnitude of response is dependent upon the surface forces as well as the dielectric properties of the liquid, thus in water even at room temperature the force of the reaction field is much less pronounced and thus harder to evaluate with this method. Surface polarization spreads along the entire surface of the liquid, and it is interesting that in the case of water a region of liquid accumulates in a band that has an apparently lower refractive index (density). This band is visible in the phase images as a blue colored feature that lies in a stable position a few millimeters beneath the surface. The region above it (that is just below the air-water interface) subsequently appears to have a normal number density. This low density band feature is slow to develop and only becomes pronounced several seconds after the field is activated.

The low density band is interrupted in the immediate vicinity of the needle electrode. Here we see that within a very short time (<1 sec) after the field is switched on, a higher refractive index (density) region develops, quite diffuse at first and steadily becoming more pronounced and localized close to the point electrode. The density continues to increase and the region affected shrinks in size until this more dense fluid begins to sink and forms a downward moving jet of liquid. It is a natural assumption to think that this liquid must be denser because it is colder, and indeed this is the case. When the fiber optic temperature probe was placed into the liquid beneath the needle a temperature difference of ~1°C was measured. It is interesting to note that moving the probe further away from the needle did not recover any areas were the water was warmer as one might think to be the case for the band of apparently less dense water just below the surface. So it appears that while temperature does play a role in the phenomenon it is not the only parameter being affected by the field.

It should be noted that the change in the electric field plays a role in the observations. Switching the electric field on or off yields the most dramatic response and with regards to the jets the change in electric field state (i.e. on or off) elicits the same response. In the steady state, the dynamics are far more irregular with jets coming and going, and is likely due to instabilities in the system caused by temperature driven dynamics or ionic wind along the surface. There is a clear difference, however, between the jets which form during a change in the field and those that are present during steady state conditions. The later tend to move slowly back and forth along the surface at a rate of ~0.2 – 0.5 mm/s whereas the jets produced by the voltage step form directly below the needle and do not drift along the surface but rather propagate downward ~0.8 mm/s and dissipate within 10 seconds. Additionally, the jets present during steady state conditions have a weaker effect on the optical phase rarely exceeding +3π radians compared the +6π radians shift when the field change is applied.

*3.2 Mid-Infrared Raman Imaging – polarizability and extinction ratio*

In a similar manner to interferometry Raman spectroscopy also probes the electric polarization of molecules - the difference is in the spatial resolution. Unlike the macroscopic refractive index interrogated by interferometry the Raman bands are a probe of the molecular transition dipole population distribution in the focal region of the laser beam. The connection between the vibrational line shape and molecular environment which make up these bands is an ongoing effort [30,31] and like most spectral features in condensed matter is obscured by inhomogeneous broadening [32,33]. The vibrational modes of water are both infrared and Raman active. The shape of these bands are complex as expected for condensed matter and have been extensively discussed elsewhere [34–36] and will be summarized here. The Raman scattering spectrum provide ensemble information on the coupling between vibrational oscillators in the liquid. The isotropic spectrum has a somewhat complex shape which has been the subject of much scrutiny. Suffice it to say that those oscillators with lower wavenumber (3250 cm$^{-1}$) on the red-side of the distribution correspond to vibrational oscillators which are damped by interactions with their local environment. Those oscillators with less degrees of interaction have higher frequencies (3490 cm$^{-1}$) and the small shoulder on the blue-side of the distribution at 3650 cm$^{-1}$ has been shown through computer simulation using experimental data as fit to belong to free oscillators which behave as single-particles.[37] The lifetime of such free particles is considered to be quite short lived.

The polarized Raman spectrum is obtained when the analyzer and incident electric field are aligned. Similarly when the analyzer is rotated perpendicular to E$_0$ the anisotropic depolarized spectrum is recorded. The isotropic spectrum can be recovered from the measured spectra via:

$$I_{iso}(\omega) = I_{VV}(\omega) - \frac{4}{3}I_{VH}(\omega) \qquad (1)$$

Whereas, the anisotropic spectrum is contained entirely in the depolarized spectrum $I_{VH}(\omega)$.

The Raman active bands in water can be attributed to a number of physical quantities, fundamental to all is the electric polarizability of the molecule which is probed directly by the electric field of the off-resonant incident linearly polarized pump radiation. The decay of the induced transition dipoles results in a frequency shifted Raman emission which contains contributions from the isotropic and anisotropic polarizabilities.

Three experimental variables in addition to the scattering polarization were examined:

    1) needle position relative to the water surface

    2) laser beam orientation relative to the liquids surface, and

    3) the applied DC electric field either on or off.

The isotropic Raman spectra for each of the experimental configurations are shown in figure 5 and reveal small changes in intensity when voltage is applied. Each of the 25,600 single spectra are background subtracted, summed and then intensity normalized; providing the cumulative polarizability response of water to steady state electric field. When (5a) the needle is in air ~2 mm above the liquid surface and the laser beam parallel the least amount of change in the spectrum is observed. When the needle is subsequently submerged 4 mm below the liquid surface (5b) so that the laser beam is now probing the area 1 mm deeper than the needle the intensity increases most. Submerging the needle further (5c) to a depth of 6 mm – 1 mm deeper than the laser beam – reduces the scattering intensity of the central frequencies. This contrasts the response when the laser beam is perpendicular to the liquid surface (5d) and the needle is submerged 4 mm now there is no observable change in the central frequencies but rather enhancement in both wings of the isotropic spectrum as a result of the applied voltage.

In order to recover changes in the frequency distributions of the probed chromophores in response to the applied field, the absolute difference spectrum (grey solid line) was calculated and plotted on the second y-axis in figure 5a-5d. . In both panels (5a) and (5c) the shape is similar and is what would be expected given a small shift in temperature[38]. The peak frequency of the difference spectrum shifts more towards the blue side (5a) when the needle is in the air indicative of heating, whereas the red shift when the needle is submerged (5c) fits in the case of cooling. The spectral shape of (5b) and (5d) have a bimodal distribution although (5b) is much narrower than (5d). These spectra are obtained from areas where the grad $E_a$ is the strongest. In the case of $E_0$ parallel to $E_a$ ($E_0 \parallel E_a$, 5b) the residual difference resembles a superposition of those from (5a) and (5c). Whereas when $E_0$ perpendicular to $E_a$ ($E_0 \perp E_a$ , 5d) there is a clear enhancement in both the red and blue wings of the spectrum. This situation is perplexing as it would appear to indicate that competing energy modes – cooperative and single particle vibrational relaxation – are enhanced rather than the expected behavior where the population density shifts from one mode to the other. This indicates a departure of the system from the ground equilibrium state.

In order to better ascertain the positional (and field) dependency of these observed changes the spectral imaging maps are examined in figure 6 where $E_0 \parallel E_a$ and figure 7 where $E_0 \perp E_a$. The maps are the average of 50 images filtered using an asymmetric, 2D, third order, Savitzky-Golay filter to reduce spurious noise and intensity fluctuations between adjacent spectra. The resulting maps without applied voltage are then subtracted from those with voltage for both the isotropic and anisotropic polarizability tensor and are displayed in the panels right and left, respectively, of center. The resulting maps provide an image of the difference in the molecular polarizability density of states caused by $E_a$. The displayed orientation of the maps matches that of the laser beam (and entrance slit) so that in figure 6 distance is plotted along the abscissa whereas in figure 7 it is plotted along the ordinate. This is done to reduce confusion regarding relative orientations between configurations. The experimental configuration used is illustrated in the central panel for each row. The needle electrode is the solid black line whose top location is highlighted within a circle and the height indicated by the dashed line. The red solid line indicates the laser beam location, and the grey square the area imaged, however, only the illuminated volume of the beam is sampled and additional spatial filtering is provided by the spectrograph entrance slit. The spectra are mapped over the same frequency range (2400 cm$^{-1}$ – 4200 cm$^{-1}$) as is done throughout the paper. When the beam is parallel to the surface we have the opportunity to examine the lateral dissipation of any induced polarization. The location of the needle is shown with a black dashed line. Adjacent to each spectral map is an additional panel where three spectra are displayed that represent the summed difference intensities from one of three regions on the map. The color of the line corresponds with the highlight color on the distance axis and follows the progression red, green, blue moving successively further from the needle electrode. The red curve corresponds to the region closest to the needle in all measurements.

Let us first examine the results when $E_0 \parallel E_a$ shown in figure 6. The central frequencies of the anisotropic polarizability are suppressed when the needle is in air (position I) and the voltage is applied. The response is strongest close to the needle as can be seen in the red spectrum. Once the needle is submerged (positions II and III) this response is inverted, the central frequencies are now enhanced. When the beam is nearest the needle (position II) the response is the weakest, the extracted signal is close to the noise floor, and there appears to be no spatial variance. With the needle deeper (position III) the response becomes stronger as the voltage enhances the anisotropic polarizability. Thus, central frequencies of the anisotropic response appear to be tuneable by varying location of a dense inhomogeneous electric field with respect to the liquid surface.

The isotropic polarizability is much stronger than the anisotropic and it is thus better suited for observing changes due to polarization. When the needle is in air (position I) the blue side of the spectrum shows strong enhancement similar to what is found in figure 5a and this is again strongest nearest the needle where the field density is highest. The blue curve shows a bimodal distribution enhancement furthest from the intense field. Submerging the needle has an exceptionally strong effect on the isotropic spectrum when the beam is just below the needle (position II). This is the area where the magnitude of $E_a$ is the greatest and the effects of local refractive index change may play a non-trivial role in the resulting spectral maps. Indeed the intensity is found to fall off quite quickly once the beam passes the needle's position. Potentially trivial causes of the observed spatial changes in the recovered intensity can be ascribed to either the deflection of the beam due to local refractive index changes associated with the cold jets or geometric misalignment. However, it is expected that such changes in scattering due to the propagation of the pump radiation would also be observed in the anisotropic spectrum (Fig.

6 left panels) which it is not. Additionally, the absorption cross section and expected scattering losses over the mm length scale are considered too small to play any significant role. The overall enhancement of intensity is strongest before reaching the needle and is more than 4x stronger than in the case of position I. The relative change of the mode density is more symmetric than in position I. With the needle fully submerged (position III) the isotropic modes are now suppressed furthest from the needle and weakly enhanced closes to the needle. The spatial image is also quite intriguing as nearest the needle the intensity difference becomes weakly positive. This may be due to the reflection of scattered radiation back towards the detector by the needle which was placed behind the laser beam. The red curve shows what would be expected for a decrease in temperature – namely the enhancement of the low frequency collective mode and loss of high frequency single-particle modes.

Rotating the laser beam perpendicular to the surface also changes the orientation of the probing electric field, the results are shown in figure 7. The laser beam was focused next to the needle electrode and the region of interest positioned so that the needle was less than 1mm below the top of the frame thus minimal optical interaction is expected with the needle and the scattered radiation. In both the anisotropic and isotropic spectral maps there is a clear effect of the electric field on the scattering intensity. The greatest influence is found where the inhomogeneous field gradient is the strongest. While the central frequencies of the isotropic spectra are clearly depleted by the field, those oscillators with frequencies at the extremes of the bandwidth are enhanced in agreement with figure 5d. This observation brings back the question of a bimodal distribution in the vibrational band structure of water under the influence of an inhomogeneous field. An estimate of population size, taken from the integrated intensity values, that is affected by the field shows this number to be quite small ~1% of the probe volume which is corresponds to approximately $7 \times 10^{16}$ oscillators.

The influence of temperature on the Raman spectra of water in the presence of inhomogeneous electric field gradients bears further investigation, however, direct measurement of the region directly beneath the needle using the fiber optic temperature probe found the temperature decreased ~0.3°C when the inhomogeneous field was applied to 20°C water.

In summary, the applied inhomogeneous electric field redistributes the vibrational population of ambient water. The effect is dependent upon the magnitude of the field gradient. The polarization response tends toward a bimodal distribution where the gradient is strongest. Where the field is weaker temperature effects become apparent in the spectra. This is consistent with the anatomy of the flow dynamics observed from interferometry. A locally focused region of cooler liquid confined below the needle electrode moves downward and dissipates several tens of mm below the surface.

*3.3 The effect of exciting a delocalized vibrational state*

Auer and Skinner [39] examined the source of the collective modes which give rise to the band at 3250 cm$^{-1}$ and found this oscillator frequency is due to the delocalization of excitation energy over as many as 12 chromophores (i.e. individual vibrational oscillators). Later research also with Skinner's group [32] extended this finding to be the result of the *"delicate quantum interference effects"* driven by local excited states which spatially extend the eigenstates and thus redistribute vibrational energy within the total population. In other words a small number of local excitations can reshape the frequency distribution of the oscillators. These excitations are not stationary and evolve in the motional limit of the liquid. An extended or coherent oscillator population will be observed as an intensity increase relative to the normal total population size. Such an increase is observed in the fluid volume where the inhomogeneous field gradient is strongest – that is, closest to the needle. It is reasonable to consider that this population is the direct result of dielectric polarization. The polarization of the water establishes a dielectric permittivity gradient which further focuses the electric field by the action of dielectric saturation [40] whereby the effective electric field strength is increased due to space charge effects induced by the dielectric medium.[41,42] The observed changes in the Raman cross section (Fig. 7) are reflective of the development of this space charge and can be thought to be similar to the onset of charge density wave effects observed in solid state materials[43,44]. The strong degenerate nature of the OH stretch transition frequencies in water means that the vibrational coupling in water is exceptionally strong.[37] This strongly mixed nature of inter- and intra-molecular vibrations in liquid water limits the use of traditional relaxation models when attempting to derive a molecular level interpretation of the resulting shifts in the Raman spectra[45].

Regardless of such limitations, the observed Raman spectra show enhanced intensities under certain conditions, i.e. relative position of the laser beam with respect to the point electrode and beam polarization, which indicates an increased Raman cross section of the observed modes, i.e. an enhanced polarizability change associated with the underlying vibration. Different effects are observed in different parts of the OH-stretch vibrational band in the observed Stoke shift frequency window. The vibrational frequency of intramolecular OH stretch vibrations of water molecules is tightly linked to the connectivity within the local hydrogen bond network[34,37,46], which also affects other properties such as the collective or localized character of the vibrational modes, i.e. the distribution of vibrational motion over adjacent molecules[47,48]. The Polarization of water due to preferential orientation, as well electronic polarization within the molecules within the applied fields employed in this work, are likely responsible for the local changes in the observed vibrational spectra. Modified hydrogen bonding motifs in water with preferentially aligned water molecules are expected to alter vibrational frequencies, while electronic polarization effects will modulate the observed intensities. The stability of local hydrogen bonding motifs has been found to be susceptible to a polarized electric field[49,50] however, the applied field strengths are three orders of magnitude greater than those employed in this work. In addition the expected preferential molecular orientations appear to be quite weak as evident from 2D neutron diffraction studies of EHD liquid bridges.[11] Thus, we consider that the field gradient rather than the total field magnitude is responsible for the observed changes in the Raman cross-section. The gradient produces a local distortion in the molecular polarizability strong enough that previously forbidden[51] or higher energy transitions become accessible[52]. The emergence of a proton channel and the stable generation of a delocalized proton state in the water molecule previously observed in EHD liquid bridges[15,53], again become viable mechanisms whereby a physical body couples to and anchors the collective vibrational mode[54].

A small excited state population is stabilized against thermal perturbation because the field gradient introduces an asymmetric polarization of molecular dipoles which establishes a unidirectional heat flow. By the mechanisms discussed above it is expected that once this population reaches a threshold density the delocalized collective oscillations overlap and this will change the overall dynamic structure of the liquid. Hydrogen bonding will be reinforced as those molecular dipoles that lie along the field gradient lines will be pinned. The population size of pinned molecules is relative to the local temperature and the magnitude of the field gradient. These pinned molecules will anchor hindered rovibronic motions (i.e. librations) which retard the thermalization of absorbed vibrational energy to the hydrogen bond (HB) network or of surrounding molecules or bath. The spectral signature for such the OH librational mode of water at 2560cm$^{-1}$ has been observed in floating water bridges with an intensity above the thermal blackbody radiation distribution.[15] This type of non-equilibrium population will similarly alter the physical properties of the liquid and increase the enthalpy due to the retardation of energy dissipation in the network. The trapping of energy in local modes would increase the number of oscillators at 3490 cm$^{-1}$ and this is seen in the Raman spectra nearest the needle. The observation that the temperature decreases slightly but the local energy increases now becomes understandable. Furthermore, the local entropy will also be affected in such a way as to promote local ordering which further reinforces the fine quantum coupling responsible for the collective mode further extending the coherence length of local vibrational states.

Ultrafast vibrational spectroscopy floating water bridges [55] supports this view of an excited molecular population with enhanced collective modes. The vibrational lifetime for the OH stretch of HDO in the bridge at 25 °C (630±50 fs) lies between the lifetimes found in bulk water at 0 °C (630±50 fs) and ice Ih at 0 °C (483±16 fs) [56] indicating that intermolecular coupling is enhanced. The dissipation of the absorbed energy to the bath however, is six fold slower in the bridge than bulk HDO:$D_2O$. Taking these relaxation times with respect to the thermodynamic limit we find that the enthalpy of water in the floating water bridge is increased while the local entropy is decreased. This fits with the model discussed above and is further supported by the Raman spectra presented here.

In contrast with electrically driven equilibrium phase transitions reported and discussed elsewhere in the literature,[57–59] the behavior elucidated here is best understood as the emergence of metaproperties dependent upon an excited state. The inhomogeneous field gradient holds the liquid in a steady-state far from equilibrium with the environment. If the field is extinguished the population collapses as is expected for dissipative structures that obey non-equilibrium thermodynamics and exhibit non-reversible behavior.[60,61] The generation of a vibronically coupled "pinned" molecular population is a non-adiabatic process, there is heat flow in order to balance entropy production; changing the bath entropy which will affect the flow of heat into or out of the region. This is essentially a thermoelectric effect governed by the Onsager reciprocal relations and which is responsible for establishing the electro-convective flow [40,62] observed here interferometrically. There exists a reciprocal relationship between electric and thermal field gradients in water [63,64] and thus the presence of one will establish the other and in a sense provide a stabilizing influence. Mass transport out of the locally polarized region, driven both by the gradients in density and chemical potential is further accelerated by electrokinetic forces arising at interfaces in the system.[65] In addition to the air-water interface an internal dynamical surface is also present. This surface is defined by how far the electric field gradient extends in the liquids with a magnitude greater than the dissipative energies of the bulk which are proportional to both the temperature and any shear forces present. The de-excitation of molecules is thus driven by electroconvective transport which depletes the population and establishes an upper bound on the size. It is expected that this would be a materially dependent feature and different polar liquids should exhibit differences in this behavior.

3.4 Implications of an extended vibrational energy mode

From the observations we find evidence to suggest that in liquid water an electric field gradient which is sufficiently steep to polarize the molecular dipoles and to affect the electronic and protonic distributions is also sufficient to cause the decoupling of local and collective vibrational modes over length scales greater than immediate nearest neighbors. This finding shows an inherent ability of liquid water to support vibrational demixing at the mesoscale.[66] Such behavior has already been noted for the mechanism of ultrafast energy relaxation whereby the local excited intramolecular vibrational modes rapidly couple to intermolecular vibrational and bending modes.[34,45] The energy transfer proceeds as the excitation traverses configurations with ever lower energy level until finally the absorbed perturbation becomes trapped in a strongly hydrogen bonded trap.[45] The important distinction is that the relaxation pathway in water is one which utilizes the strong interaction of the hydrogen bond, which in the case of the experiments presented here exhibit enhanced polarization [67,68]. When one expands the number of molecules involved in the relaxation process a length scale emerges whereby the local and collective modes can effectively decouple from each other – the mesoscale [24,69,70]. In the presence of the field gradient this decoupling is enhanced and essentially results in transiently isolated molecular populations within the liquid. Given that such electric field gradients as those found here have also been observed in a number of natural systems, e.g. cells[26], clay particles[71], and cloud droplets[72], it is possible to consider that within a volume of physically continuous liquid isolated regions with unique properties can form. The interplay between the reactive force generated by the polarization of water dipoles and the driving field gradient establishes a stabilizing dis-equilibrium that is spatially defined by the magnitude of the applied field gradient and the dipole response dynamics of the molecules. The electric field gradients are thus sufficient to stabilize local regions whereby the chemical potential follows the gradient, and thus reactivity can be spatially confined without the need for a physical barrier such as a membrane. The high charge density, and well defined hydrophobic/hydrophilic profiles, of many biomolecules provides the ideal situation to maintain such mesoscale pockets of augmented water activity. The asymmetric transport of energy in liquid water can thus be effectively controlled by the presence of strong electric field gradients and likely plays a role in water regulation of living cells.[73]

## 4. Conclusions

The thermoelectric effect is observed in water using interferometry and mid-IR Raman spectroscopy. While the interferometry data reveals macroscopic changes in the polarizability of the liquid that occur alongside changes in other observables such as density and temperature, the spectroscopic information reveals unexpected changes at the microscale. The vibrational relaxation of water molecules is strongly governed by intermolecular interactions and these interactions do correlate with temperature, however, the expected behavior is not observed in the case where a strong inhomogeneous electric field gradient is present. Rather, the difference in Raman scattering spectra collected with and without field gradient show enhancement of collective vibrational modes as well as an increase in total system energy. This indicates the emergence of a bimodal distribution in the transient dipole population stabilized against thermal equilibrium by the inhomogeneous electric field. This has the effect of inducing directionality in the flow of heat through the system and establishes the microscopic origin for electrohydrodynamic flow.

**Acknowledgements**

This work was performed in the cooperation framework of Wetsus, European Center of Excellence for Sustainable Water Technology (www.wetsus.eu). Wetsus is co-funded by the Dutch Ministry of Economic Affairs and Ministry of Infrastructure and Environment, the Province of Fryslân, and the Northern Netherlands Provinces. The authors wish to thank the participants of the research theme Applied Water Physics for the fruitful discussions and their financial support.

**References**

1  W. Armstrong, *Account of the transactions of the Newsactle Literary and Philosophical Society*, 1893.

2  K. Morawetz, *Phys. Rev. E - Stat. Nonlinear, Soft Matter Phys.*, 2012, **86**, 1–9.

3  A. G. Marín and D. Lohse, *Phys. Fluids*, 2010, **22**, 122104.

4  R. Moro, J. Bulthuis, J. Heinrich and V. V. Kresin, *Phys. Rev. A - At. Mol. Opt. Phys.*, 2007, **75**, 1–6.

5  C. M. Whitehouse, R. N. Dreyer, M. Yamashita and J. B. Fenn, *Anal. Chem.*, 1985, **57**, 675–679.

6  G. D. Martin and I. M. Hutchings, in *Inkjet Technology for Digital Fabrication*, eds. I. M. Hutchings and G. D. Martin, John Wiley & Sons, Ltd, Chichester, UK, 2012.

7  J. H. Wendorff, S. Agarwal and A. Greiner, *Electrospinning*, Wiley-VCH Verlag GmbH & Co. KGaA, Weinheim, Germany, 2012.

8  L. L. F. Agostinho, PhD Thesis, Technical Univeristy Delft, Netherlands, 2013.

9  L. B. Skinner, C. J. Benmore, B. Shyam, J. K. R. Weber and J. B. Parise, *Proc. Natl. Acad. Sci. U. S. A.*, 2012, **109**, 16463–8.

10  E. C. Fuchs, B. Bitschnau, J. Woisetschläger, E. Maier, B. Beuneu and J. Teixeira, *J. Phys. D. Appl. Phys.*, 2009, **42**, 065502.

11  E. C. Fuchs, P. Baroni, B. Bitschnau and L. Noirez, *J. Phys. D. Appl. Phys.*, 2010, **43**, 105502.

12    O. Fuchs, M. Zharnikov, L. Weinhardt, M. Blum, M. Weigand, Y. Zubavichus, M. Bär, F. Maier, J. D. Denlinger, C. Heske, M. Grunze and E. Umbach, *Phys. Rev. Lett.*, 2008, **100**, 249802.

13    E. C. Fuchs, B. Bitschnau, S. Di Fonzo, A. Gessini and J. Woisetschläger, 2011, **1**, 135–147.

14    L. Piatkowski, A. D. Wexler, E. C. Fuchs, H. Schoenmaker and H. J. Bakker, *Phys. Chem. Chem. Phys.*, 2012, **14**, 6160–4.

15    E. C. Fuchs, A. Cherukupally, A. H. Paulitsch-Fuchs, L. L. F. Agostinho, A. D. Wexler, J. Woisetschläger and F. T. Freund, *J. Phys. D. Appl. Phys.*, 2012, **45**, 475401.

16    R. C. Ponterio, M. Pochylski, F. Aliotta, C. Vasi, M. E. Fontanella and F. Saija, *J. Phys. D. Appl. Phys.*, 2010, **43**, 175405.

17    M. Sammer, A. Wexler, P. Kuntke, H. Wiltsche, N. Stanulewicz, E. Lankmayr, J. Woisetschlaeger and E. Fuchs, *J. Phys. D. Appl. Phys.*, 2015.

18    A. Widom, J. Swain, J. Silverberg, S. Sivasubramanian and Y. Srivastava, *Phys. Rev. E*, 2009, **80**, 016301.

19    J. Woisetschläger, A. D. Wexler, G. Holler, M. Eisenhut, K. Gatterer and E. C. Fuchs, *Exp. Fluids*, 2012, **52**, 193–205.

20    V. I. Arkhipov and N. Agmon, *Isr. J. Chem.*, 2003, **43**, 363–371.

21    F. Kremer and A. Schönhals, *Broadband Dielectric Spectroscopy*, 2003.

22    C. A Stan, S. K. Y. Tang, K. J. M. Bishop and G. M. Whitesides, 2011, 1089–1097.

23    W. Stygar, T. Wagoner, H. Ives, Z. Wallace, V. Anaya, J. Corley, M. Cuneo, H. Harjes, J. Lott, G. Mowrer, E. Puetz, T. Thompson, S. Tripp, J. VanDevender and J. Woodworth, *Phys. Rev. Spec. Top. - Accel. Beams*, 2006, **9**, 070401.

24    H. Tanaka, *Faraday Discuss.*, 2013, **167**, 9.

25    V. Raicu and Y. Feldman, *Dielectric Relaxation in Biological Systems Physical Principles*, Oxford, 2015.

26    K. M. Tyner, R. Kopelman and M. a Philbert, *Biophys. J.*, 2007, **93**, 1163–74.

27    R. G. Harrison, K. L. Aplin, F. Leblanc and Y. Yair, *Planetary atmospheric electricity*, 2008, vol. 137.

28    A. G. Cairns-Smith, *Genetic Takeover: And the Mineral Origins of Life*, 1987.

29    M. Hipp, J. Woisetschläger, P. Reiterer and T. Neger, *Measurement*, 2004, **36**, 53–66.

30    L. Shi, Y. Ni, S. E. P. Drews and J. L. Skinner, *J. Chem. Phys.*, 2014, **141**, 084508.

31    C. J. Tainter, L. Shi and J. L. Skinner, *J. Chem. Theory Comput.*, 2015, **11**, 2268–2277.

32    M. Yang and J. L. Skinner, *Phys. Chem. Chem. Phys.*, 2010, **12**, 982–991.

33    L. Shi, S. M. Gruenbaum and J. L. Skinner, *J. Phys. Chem. B*, 2012, **116**, 13821–30.

34    H. J. Bakker and J. L. Skinner, *Chem. Rev.*, 2010, **110**, 1498–517.

35    G. E. Walrafen, *J. Chem. Phys.*, 2005, **122**, 174502.

36   B. M. Auer and J. L. Skinner, *Chem. Phys. Lett.*, 2009, **470**, 13–20.

37   J. Skinner, B. Auer and Y. Lin, *Adv. Chem. Phys.*, 2010, **142**.

38   H. Torii, *J. Phys. Chem. A*, 2006, **110**, 9469–9477.

39   B. M. Auer and J. L. Skinner, *J. Chem. Phys.*, 2008, **128**, 224511.

40   U. Kaatze, *J. Solution Chem.*, 1996, **26**, 1049–1112.

41   K. Adamiak and P. Atten, *J. Electrostat.*, 2004, **61**, 85–98.

42   A. Kara, Ö. Kalenderli and K. Mardikyan, *COMSOL Conf. 2006*, 2006.

43   S. Sugai, *Phys. Rev. B*, 1984, **29**, 953–965.

44   P. Goli, J. Khan, D. Wickramaratne, R. K. Lake and A. a Balandin, *Cond-mat*, 2012, 1211.4486.

45   K. Ramasesha, L. De Marco, A. Mandal and A. Tokmakoff, *Nat. Chem.*, 2013, **5**, 935–40.

46   Y. Maréchal, *J. Mol. Struct.*, 2011, **1004**, 146–155.

47   R. Ludwig, *Angew. Chemie - Int. Ed.*, 2001, **40**, 1808–1827.

48   D. E. Moilanen, E. E. Fenn, Y. Lin, J. L. Skinner, B. Bagchi and M. D. Fayer, 2008, **105**.

49   D. Rai, A. D. Kulkarni, S. P. Gejji and R. K. Pathak, *J. Chem. Phys.*, 2008, **128**, 034310.

50   D. Rai, PhD Thesis, University of Pune, Pune, India, 2008.

51   E. Ayars, H. Hallen and C. Jahncke, *Phys. Rev. Lett.*, 2000, **85**, 4180–4183.

52   H. J. Bakker and H.-K. Nienhuys, *Science (80-. ).*, 2015, **297**, 587–590.

53   E. C. Fuchs, B. Bitschnau, A. D. Wexler and F. Woisetschläger, Jakob Freund, *J. Phys. Chem. B*, 2015.

54   R. L. A. Timmer, K. J. Tielrooij and H. J. Bakker, *J. Chem. Phys.*, 2010, **132**, 194504.

55   L. Piatkowski, A. D. Wexler, E. C. Fuchs, H. Schoenmaker and H. J. Bakker, *Phys. Chem. Chem. Phys.*, 2012, **14**, 6160–4.

56   S. Woutersen, U. Emmerichs, H.-K. Nienhuys and H. Bakker, *Phys. Rev. Lett.*, 1998, **81**, 1106–1109.

57   E.-M. Choi, Y.-H. Yoon, S. Lee and H. Kang, *Phys. Rev. Lett.*, 2005, **95**, 085701.

58   J. Bartlett, A. van den Heuval and B. Mason, *Zeitschrift für Angew. Math. und Phys. ZAMP*, 1963, **14**, 599–610.

59   I. M. Svishchev and P. G. Kusalik, *J. Am. Chem. Soc.*, 1996, **118**, 649–654.

60   I. Prigogine, *Science (80-. ).*, 1978, **201**, 1065–1071.

61   A. Kleidon and R. D. Lorenz, Eds., *Non-equilibrium Thermodynamics and the Production of Entropy*, Springer-Verlag Berlin Heidelberg, 2005.

62   J. B. Hubbard, L. Onsager, W. M. van Beek and M. Mandel, *Proc. Natl. Acad. Sci. U. S. A.*, 1977, **74**, 401–4.

63   W. Evans, J. Fish and P. Keblinski, *J. Chem. Phys.*, 2007, **126**, 126–129.

64  F. Bresme, A. Lervik, D. Bedeaux and S. Kjelstrup, *Phys. Rev. Lett.*, 2008, **101**, 2–5.

65  J. R. Melcher and G. I. Taylor, *Annu. Rev. Fluid Mech.*, 1969, **1**, 111–146.

66  R. E. Van Vliet, M. W. Dreischor, H. C. J. Hoefsloot and P. D. Iedema, *Fluid Phase Equilib.*, 2002, **201**, 67–78.

67  S. Woutersen and H. J. Bakker, *Nature*, 1999, **402**, 507–509.

68  E. T. J. Nibbering and T. Elsaesser, *Chem. Rev.*, 2004, **104**, 1887–1914.

69  P. K. Gupta and M. Meuwly, *Faraday Discuss.*, 2013, **167**, 329.

70  T. Yagasaki, M. Matsumoto and H. Tanaka, *Phys. Rev. E - Stat. Nonlinear, Soft Matter Phys.*, 2014, **89**, 1–5.

71  S. Yariv and H. Cross, *Geochemistry of Colloid Systems: For Earth Scientists*, Springer Science & Business Media, 2012.

72  D. R. MacGorman and W. D. Rust, *The Electrical Nature of Storms*, 1998.

73  B. L. De Groot, T. Frigato, V. Helms and H. Grubmüller, *J. Mol. Biol.*, 2003, **333**, 279–293.

Figures

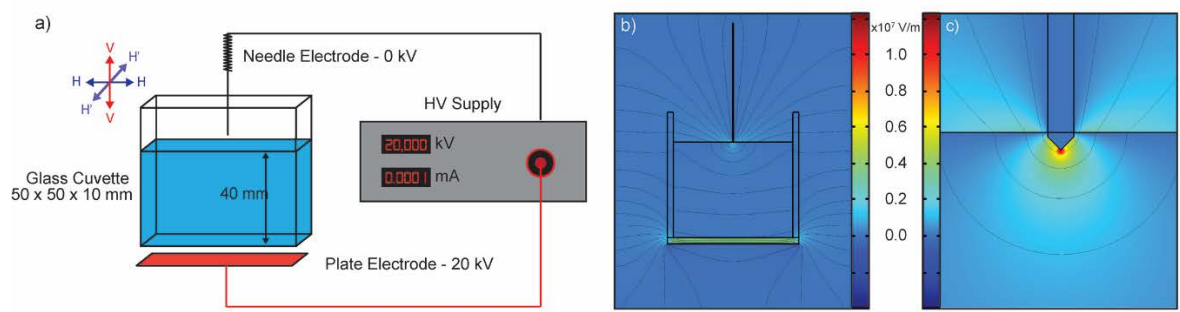

Figure 1 Experimental configuration used in this study. The point plane electrode and liquid cuvette (panel a) is modeled to show (b) the electric field (x$10^7$ V m$^{-1}$) and equipotential lines (1000 V per contour) and (c) the localized electric field (x$10^7$ V m$^{-1}$) strength just beneath the ground (needle) electrode. The electric field polarizations referred to in the Raman section are shown in the upper left of panel (a).

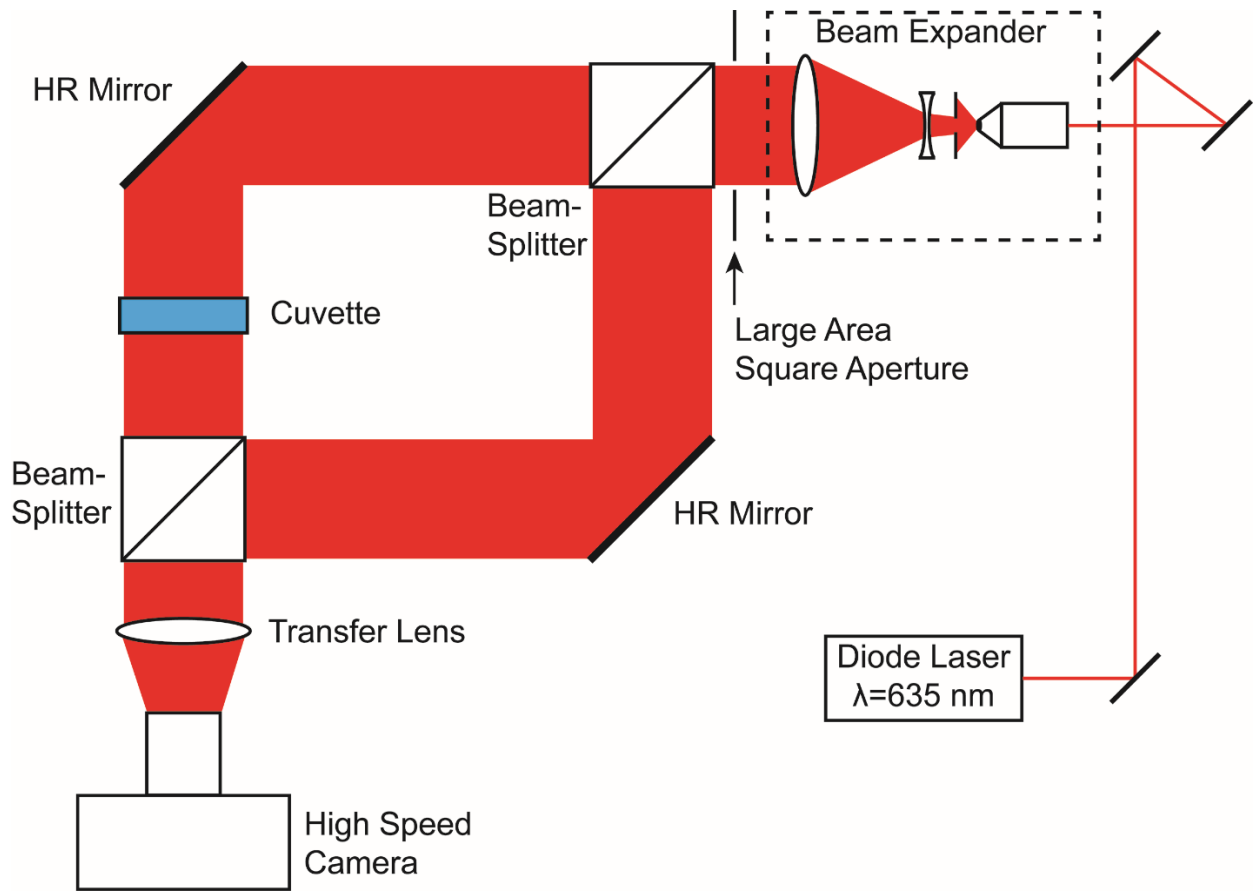

Figure 2 Diagram of the Mach-Zehnder interferometer. 2 m base length, 50 mm2 clear aperture. The beam expander consisted of a 20x objective, 5 μm pinhole, and several lenses to further diverge and collimate the spatially filtered beam.

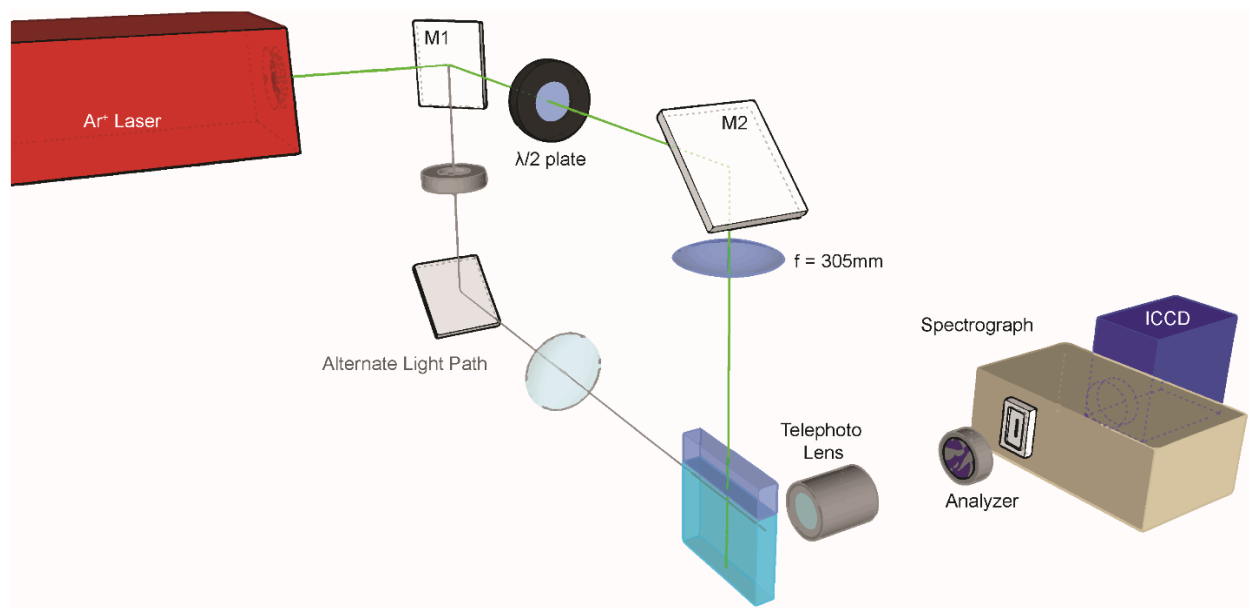

Figure 3 Diagram of the Imaging Raman spectrometer used in this study. Both light paths used are shown.

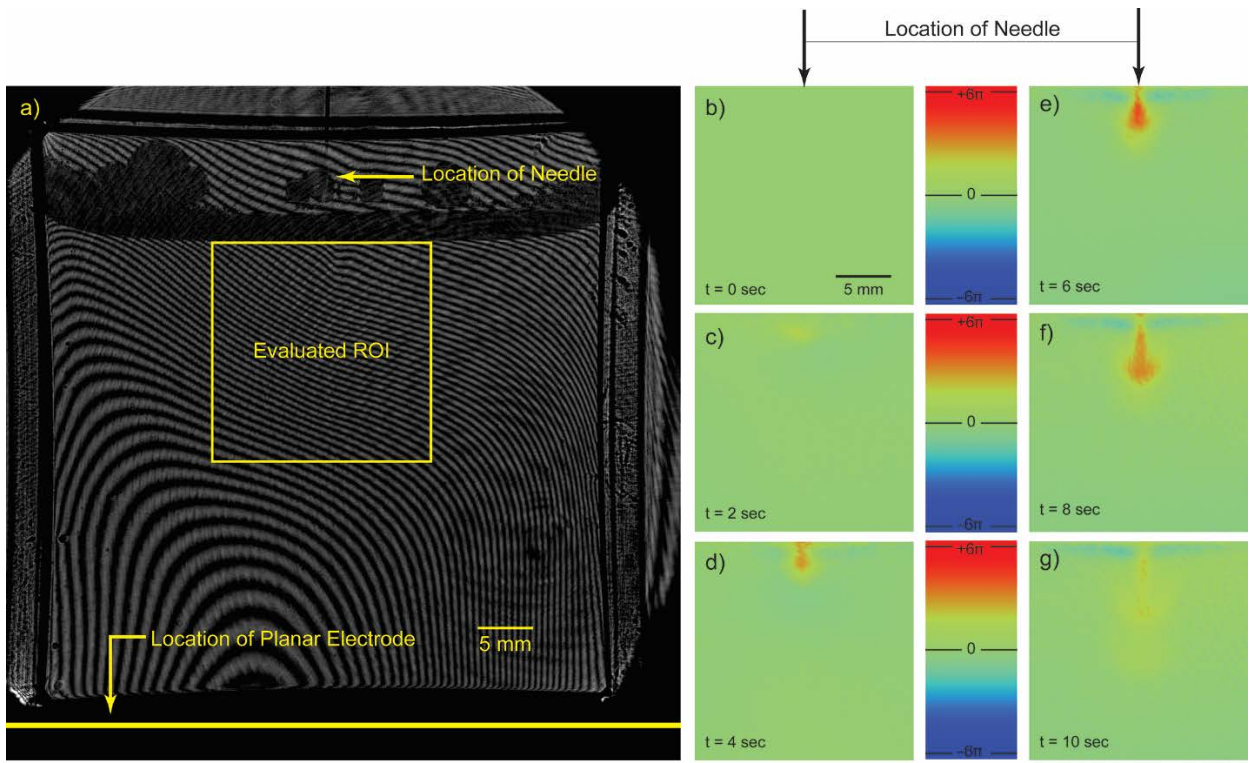

Figure 4 Interferogram of the water filled cuvette and point plane system showing the locations of electrodes and evaluated region of interest (a). Time evolution (panels b-g) for the response of the liquid to $\nabla E$ from the evaluated interferograms shows the formation of a dense downward jet of liquid constrained between regions of less dense water just below the surface.

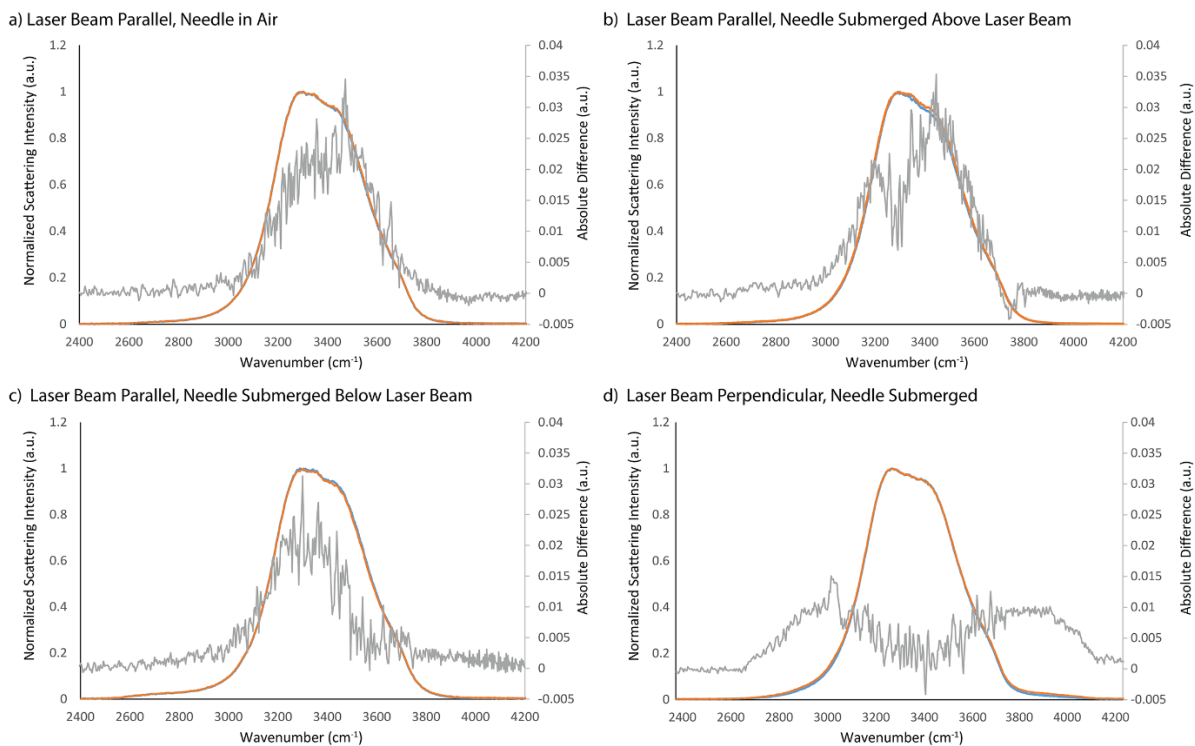

Figure 5 Normalized isotropic Raman spectra for the different measurement cases. The laser beam was oriented either parallel to the liquid surface (a-c) or perpendicular (d). In the parallel case, the laser beam was 5 mm below the surface. The needle depth was also varied so as to be (a) 2mm above the water surface, (b,d) 4 mm or (c) 6 mm below the water surface. The spectrum collected when 20kV is applied (orange line) and when no voltage is applied (blue line) are subtracted to yield the absolute difference spectrum (grey line) . The normalized scattering intensity and the absolute difference is given by the left and right ordinate scales respectively.

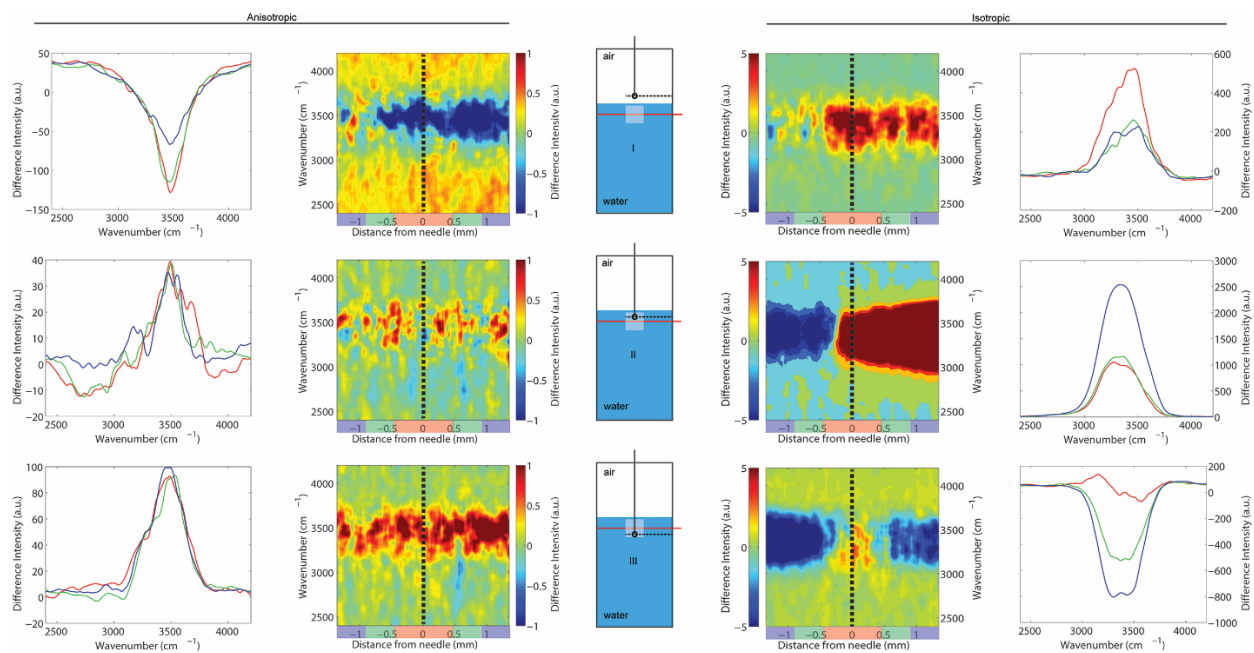

Figure 6 Raman spectral imaging with the laser beam parallel to the liquid surface. The center diagram shows the orientation of the needle (black line), the laser beam (red line), and the imaged region of interest (grey square). Three different needle depths are examined and move progressively deeper with each row. Difference spectral maps obtained by subtracting the spectra with (20 kV) and without (0 kV) electric field with the beam parallel to the liquid surface. The anisotropic map (left) and isotropic map (right) have different intensity scales. Three equal area segments from the maps are binned together to produce the line plots and are show as a shaded overlay on the abscissa of the 2D plots. The region closest to the needle corresponding to the red spectrum, the region flanking this to the left and right in green, and finally the region furthest from the needle in blue.

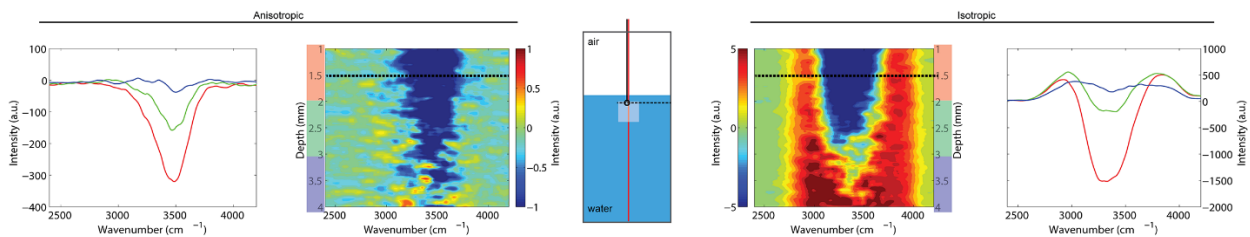

Figure 7 Raman spectral imaging with the laser beam perpendicular to the liquid surface. The center diagram shows the orientation of the needle (black line), the laser beam (red line), and the imaged region of interest (grey square). Difference spectral maps obtained by subtracting the spectra with (20 kV) and without (0 kV) electric field with the beam parallel to the needle. The anisotropic map (left) and isotropic map (right) have different intensity scales. Three equal area segments from the maps are binned together, according to the color overlay on the ordinate, and give the extracted spectra shown in the line plots. Red curves are the shallowest and closest to the needle, blue the deepest and furthest from the electrode.